\documentclass[twocolumn,showpacs,prl]{revtex4}

\usepackage{epsfig}
\usepackage{amsmath}
\usepackage{graphicx}
\usepackage{dcolumn}
\usepackage{amsmath}
\usepackage{latexsym}

\begin{document}

\title{Absence of weak antilocalization in ferromagnetic films
}
\author{N. Kurzweil}
\author{E. Kogan}
\author{A. Frydman}
\address{The Department of Physics, Bar Ilan University, Ramat Gan 52900, Israel}

\begin{abstract}
We present magnetoresistance measurements performed on ultrathin
films of amorphous Ni and Fe. In these films the Curie temperature
drops to zero at small thickness, making it possible to study the
effect of ferromagnetism on localization.  We find that
non-ferromagnetic films are characterized by positive
magnetoresistance. This is interpreted as resulting from weak
antilocalization due to strong Bychkov-Rashba spin orbit scattering.
As the films become ferromagnetic the magnetoresistance changes sign
and becomes negative. We analyze our data to identify the individual
contributions of weak localization, weak antilocalization and
anisotropic magnetoresistance and conclude that the magnetic order
suppresses the influence of spin-orbit effects on localization
phenomena in agreement with theoretical predictions.
\end{abstract}

\date{\today}

\maketitle
\section{ \textrm{I. Introduction}}

Ferromagnetic thin films are of great interest because of their
application in various areas of magnetoelectronics, magnetic
read-heads, field sensors, random access memory elements and others.
Most of the research effort is concentrated on ultraclean films which are
advantageous for devices. However, disordered films are
interesting from the scientific point of view since they allow to study the interplay between
magnetization and localization. Like  in
normal metals, the semi-classical Drude conductivity at low
temperatures  is expected to be strongly influenced by quantum
corrections.

Quantum corrections to electrical conductivity of normal metals
were extensively studied in the past three decades. These
corrections, derived from quantum interference between
self-intersecting paths \cite{localization}, are always negative in
the case of purely potential scattering. When in addition to the
potential scattering there exists spin-orbit (SO) scattering or
magnetic--impurity scattering (scattering due to magnetic impurities
in a paramagnetic metal) the situation becomes less definite. The
interference may be constructive or destructive depending on the
strength of the SO interaction \cite{Anderson, Hikami, Maekawa}. In
the general case the conductivity correction $\delta\sigma$ can be
presented as \cite{rammer}
\begin{eqnarray}
\label{ra}
&&\delta\sigma=-\frac{2e^2D_0}{\pi\hbar L^d}\\
&&\sum_{Q}'\left(\frac{\frac{3}{2}}{D_0Q^2+\frac{4}{3\tau_{SO}}+\frac{1}{\tau_{\varphi}}}
-\frac{\frac{1}{2}}{D_0Q^2+\frac{1}{\tau_{\varphi}}}\right),\nonumber
\end{eqnarray}
where $D_0$ is the diffusion constant in $d$ dimensions, $L$ is the
size of the sample, $\tau_{SO}$ is the spin-orbit relaxation time,
the inverse of which measures the strength of the SO interaction,
and  $\tau_{\varphi}$ is the dephasing time. The first term in Eq.
(\ref{ra}) is the contribution from the triplet channel, describing
the  constructive interference and leading to weak localization
(WL).  The second term  is the contribution from the  singlet
channel, describing the  destructive interference and leading to
weak antilocalization (WAL). This becomes even more obvious when one
calculates the ${\bf Q}$ sum and obtains \cite{alt2}
\begin{eqnarray}
\label{hren} \frac{\delta\sigma_d}{\sigma_d}\sim
-\int_{\tau}^{\tau_{\varphi}}\frac{\lambda^2vdt}{a^{3-d}(D_0t)^{d/2}}\left(\frac{3}{2}e^{-t/\tau_{SO}}-\frac{1}{2}\right)
\end{eqnarray}
where $\lambda$ and $v$ are the (Fermi--surface) electron wave
length and velocity respectively, and $a$ is the film thickness or
the wire radius. For $d=2$ one obtains
\begin{eqnarray}
\label{aa}
\delta\sigma_2&\sim& \frac{e^2}{\hbar}\left(-\ln\frac{\tau_{\varphi}}{\tau}\right),\qquad\qquad\qquad \tau_{SO}\gg\tau_{\varphi}\nonumber\\
&\sim&\frac{e^2}{\hbar}\left(-\frac{3}{2}\ln\frac{\tau_{SO}}{\tau}+\frac{1}{2}\ln\frac{\tau_{\varphi}}{\tau}\right),\quad
\tau_{SO}\ll\tau_{\varphi},
\end{eqnarray}
where $\tau$ is the momentum relaxation time. As seen from Eq.
(\ref{aa}), the spin--orbit scattering reverses the sign of the
quantum correction to conductivity.

Both WL and WAL are suppressed by an applied magnetic field. An
asymptotic estimate for magnetoconductance can be found from Eq.
(\ref{hren}) taking into account that the upper limit of integration
should now be $t_H$ instead of $\tau_{\varphi}$ if
$t_H\ll\tau_{\varphi}$, where $t_H\sim l_H^2/D_0$, and $l_H=(\hbar
c/2eH)^{1/2}$ is the magnetic length. It is easy to see that in the
absence of spin--orbit scattering the magnetoresistance (MR - the inverse of the magnetoconductance) is always
negative. Substituting $D_{0}=\frac{1}{3}\frac{l^2}{\tau}$
 yields the following expression for the magnetoconductance due to WL \cite{rammer}:

\begin{eqnarray}
\label{wl} \frac{\Delta\sigma_{WL}}{\sigma(0)}
=-\frac{3}{2k_F^2al}\left\{\psi\left(\frac{1}{2}+\frac{3}{4}\frac{l_H^2}{l^2}\right)
-\psi\left(\frac{1}{2}+\frac{3}{4}\frac{l_H^2}{l_{\varphi}^2}\right)\right\}
\end{eqnarray}
where  $k_{F}$ is the Fermi wave length, $l$ is the mean free path
and $l_{\varphi}$ is the dephasing length.

The situation is more complicated for WAL when spin-orbit scattering
is important. In weak magnetic fields and under strong spin--orbit
scattering the magnetoconductance reverses its sign and is given by

\begin{eqnarray}\label{wal}
&&\frac{\Delta\sigma_{WAL}}{\sigma(0)}=-\frac{3}{4k_F^2dl}\left\{2\psi\left(\frac{1}{2}+\frac{3}{4}\frac{l_H^2}{l^2}[1+\frac{l^2}{l_{so}^2}]\right)\right.\\
&&\left.-3\psi\left(\frac{1}{2}+\frac{3}{4}\frac{l_H^2}{l^2}\left[\frac{l^2}{l_{\varphi}^2}
+\frac{4}{3}\frac{l^2}{l_{SO}^2}\right]\right)
+\psi\left(\frac{1}{2}+\frac{3}{4}\frac{l_H^2}{l_{\varphi}^2}\right)\right\}.\nonumber
\end{eqnarray}
where $l_{SO}$ is the spin-orbit interaction length. As the magnetic field increases and becomes larger than
\begin{eqnarray}
H_{SO}\simeq\frac{c\hbar}{e}\frac{1}{D_0\tau_{SO}}
\end{eqnarray}
the magnetoresistance becomes negative even in the presence of SO \cite{alt2} . Similar to paramagnetic impurity scattering, SO interaction may result from lack of inversion symmetry which is described by the Dresselhaus
term \cite{Dresselhaus} or reduced dimensionality of the system related to the Bychkov-Rashba Hamiltonian \cite{rashba}.

From Eq. (\ref{ra}) it is clear that a quantity of crucial
importance for the calculation of the quantum correction to the
conductivity is the dephasing time $\tau_{\varphi}$. As well known,
in low dimensionality conductors ($d\leq 2$) for low enough
temperatures $\tau_{\varphi}$ is determined by electron--electron
interaction involving small energy transfer,
$\omega\sim\tau_{\varphi}$, and has the form \cite{alt,alt2}
\begin{eqnarray}
\frac{1}{\tau_{\varphi}}\sim\left(\frac{T}{D_0^{d/2}\nu_d\hbar^2}\right)^{2/(4-d)}\left\{\begin{array}{cl} \ln\frac{p_F^2l a}{\hbar^2}, & d=2\\
                                                                                                        1, & d=1\end{array}\right.,
\end{eqnarray}
where $\nu_d$ is the density of states, $p_F$ is the Fermi momentum.

One may ask, how do the above quantum corrections manifest
themselves in the case where the metal is a ferromagnet. Dugaev et
al. \cite{dugaev} theoretically studied the influence of
ferromagnetism on WL and WAL phenomena. Qualitatively, the main idea
in this work is that  strong magnetic polarization in ferromagnetic materials excludes processes with the singlet
Cooperon, which are responsible for the antilocalization in
nonmagnetic materials with SO scattering. As a result, the quantum
correction to conductivity is always negative in ferromagnetic samples and
leads to negative magnetoresistance. This is due to the fact that in
ferromagnetic materials ferromagnetic $s-d$ exchange yields spin
splitting which is comparable or larger than the thermal energy or
the Landau level splitting due to magnetic field \cite{sil}. The
resulting spin polarization of the conduction electrons influences
the spin-flip scattering.

For the case of a 2D ferromagnetic film, which is relevant for our
experiment, the results depend upon the orientation of the
magnetization relative to the plane of the film. The easier is the
case of the magnetization perpendicular to the plane. In this case
the spin--flip scattering, which leads to weak antilocalization  is
totally suppressed. Thus, even in the presence  of strong SO
interaction the correction to conductivity and the magnetoresistance
are negative, hence there is only weak localization correction
\cite{dugaev,sil}. When the magnetization is in plane, spin--flip
scattering is present in the  system, and one has to include the
spin--flip processes in the Cooperon ladder. However, the overall
quantum correction to the conductivity turns out to be of WL type,
though smaller than in the case of magnetization perpendicular to
the plane.

In this paper we describe a systematic experimental work designed to
study the effect of ferromagnetism on localization phenomena. For
this we use disordered amorphous films of ferromagnetic Ni and Fe in
which we vary continuously the thickness and disorder of each
sample. These films show a drop of the Curie temperature to zero at
low film thickness \cite{our}. This allows us to study the
conductivity of a \emph{single} sample - with and without
ferromagnetism - in the temperature regime in which localization is
important. We find that in the paramagnetic phase of our ultrathin
film, the conductivity is governed by WAL as a result of strong
Bychkov-Rashba SO scattering leading to positive MR. When the samples become ferromagnetic the magnetoresistance
curves change sign and become negative, indicating that WAL is
masked by magnetic effects in agreement with Dugaev el al.

\section{ \textrm{II. Experimental}}

In order to achieve substantial localization effects one would like
to employ disordered metallic thin layers. For this reason we used
ultrathin amorphous films in which the low thickness enhances the
role of disorder and thus increases the affect of localization. The
samples studied in this research were films of Ag, Ni and Fe with
thickness varying between 0.05 and 12 nm and resistances between 1
M$\Omega$ and 100 $\Omega$. The samples were fabricated using the
technique of  "quench condensation", i.e. thermal evaporation on a
cryocooled substrate. This technique allows to deposit sequential
layers of ultrathin films and measure transport without thermally
cycling the sample or exposing it to atmosphere. This is particular
advantageous for the study of thin ferromagnetic films such as Ni or
Fe in which one would like to prevent rapid oxidation characteristic
of these materials \cite{ni1,ni2}. Prior to the cooldown, six gold
leads were evaporated on an insulating Si/SiO substrate. A
mechanical mask was used to obtain a desired hall bar geometry
enabling hall effect and 4 probe conductivity measurements on a 2 by
2 mm sample.  The substrate was then connected to the He3 pot of a
He3 fridge which was pumped to high vacuum to allow film
evaporation. An insulating underlayer of Ge or Sb was predeposited
prior to the metal evaporation while the substrate was held at T=4
K. This insulating layer wets the substrate enabling to achieve
ultrathin continuous amorphous films even at monolayer thickness
\cite{strongin,goldman}. An ultrathin metal film was then evaporated
on the cold substrate. Deposition rate and film thickness were
monitored, in situ, by a quartz-crystal and sample resistance was
measured during the growth. The evaporation was terminated at a
desired thickness and resistance allowing transport measurements at
different disorder states of the sample. The magnetotransport
measurements presented in this paper were performed in the 25 K -
300 mK temperature range. A magnetic field up to 6 T was applied
perpendicular to the substrate plain.

Previously \cite{our} we studied the conductivity versus thickness
of similar ultrathin films while driving them from strong to weak
localization. Our analysis indicated that as the film is thickened,
the microscopic properties such as  mean free path, $l$,  diffusion
constant, $D_{0}$, or dephasing length, $l_{\varphi}$ do not change.
The parameter that is mainly affected by the thickness is the
localization length $\xi$ which is reduced due to the low
dimensionality of the layer. Changing the thickness causes a
crossover from strong to weak localization via the crossover from
$\xi<l_{\varphi}$ to $\xi>l_{\varphi}$.

\begin{figure}
{\epsfxsize=3.6 in \epsffile{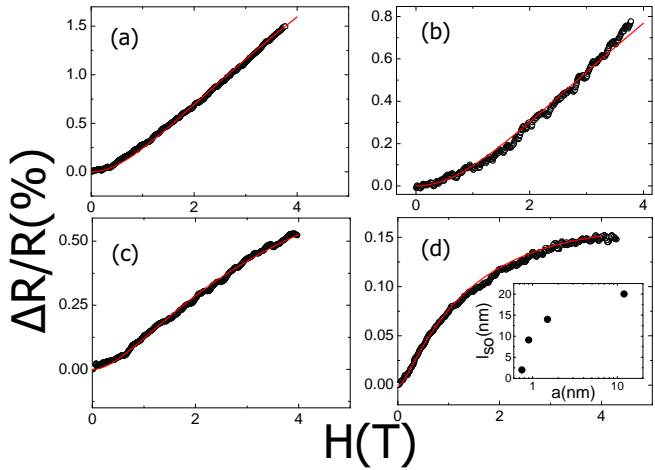}} \caption{ \small MR curves of
an Ag sample having various thickness, a. T=4.2 K. Solid lines are
fits to WAL (Eq. \ref{wal}) with mean free path of 0.2 nm. (a)
$a=0.75$ nm. (b) $a=0.9$ nm.(c) $a=1.5$ nm.(d) $a=12$ nm. The inset
shows the extracted $l_{SO}$ as a function of thickness.} \label{ag}
\end{figure}

\section{ \textrm{III. Results and discussion}}

As noted above, this research concentrates on the transport
properties of ultrathin films of Ni and Fe for the purpose of
investigating the influence of ferromagnetism on localization
effects. For reference we studied the properties of
non-ferromagnetic systems i.e. ultrathin layers of Ag. MR curves of a
single Ag sample for several thicknesses are shown in fig. 1. It is
seen that Ag layers exhibit a positive MR for all the studied
thickness range. This result is indicative of  films characterized
by  strong spin orbit interaction and thus exhibiting WAL. We
attribute these results to the presence of a strong Bychkov-Rashba
term in which SO interaction is due to scattering on the surface
caused by reduced dimensionality in the ultrathin film. The solid
lines in fig. 1 are fits to WAL (Eq. \ref{wal}). The extracted
spin-orbit length, $l_{SO}$, grows with the layer thickness (see
inset) demonstrating that the SO interaction becomes weaker as the
films become thicker, in consistency with the assumption that
Bychkov-Rashba scattering is the main contribution to the SO
effects.

\begin{figure}
{\epsfxsize=2.7 in \epsffile{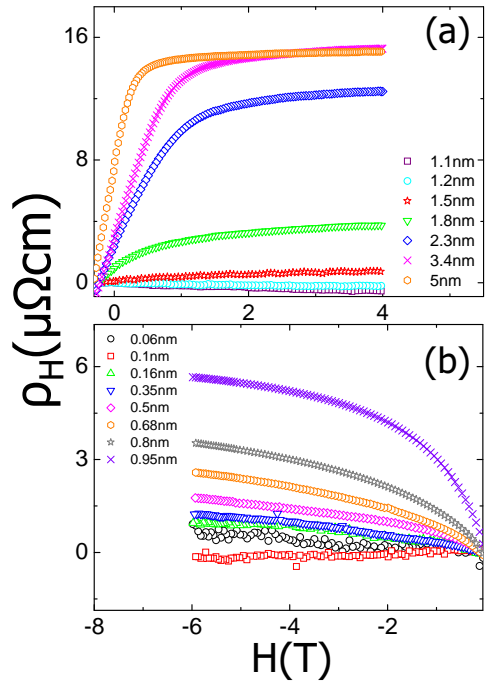}} \caption{ \small Hall Effect
measurements of sequential quench condensed Ni (a) and Fe (b) films.
T=4.2 K.}
\end{figure}

The situation is different for ferromagnetic layers. As described in
Ref. \cite{our}, thin amorphous layers of Ni or Fe show no signs of
ferromagnetism below a critical thickness, $a_{C}$, which is
material dependent. As the thickness is increased the Curie
temperature increases rapidly from zero to the bulk value. This
means that for a given temperature one can observe a transition from
paramagnetism to ferromagnetism as a function of thickness.

The measurement of magnetization in quench condensed films is
somewhat problematic since the samples can not be heated to room
temperature without affecting their geometry, hence conventional
methods such as SQUID measurements are not relevant. For this reason
we use the hall effect (HE) for which ferromagnetic films are
characterized by a large contribution from the extraordinary hall
effect (EHE) proportional to the magnetization. Fig. 2 shows Hall
effect measurements at T=4.2 K for sequential layers of Ni and Fe
films. These show that a measurable extraordinary Hall effect (EHE)
contribution, which characterizes ferromagnetism, is absent for
$a<a_{C}$, where $a_C$ is 1.8 nm for Ni and 0.45 nm for Fe. Thus we
can identify two regimes: For $a>a_C(T)$ the films are ferromagnetic
while for $a<a_C(T)$ they show no spontaneous magnetization.

\begin{figure}
{\epsfxsize=3.6 in \epsffile{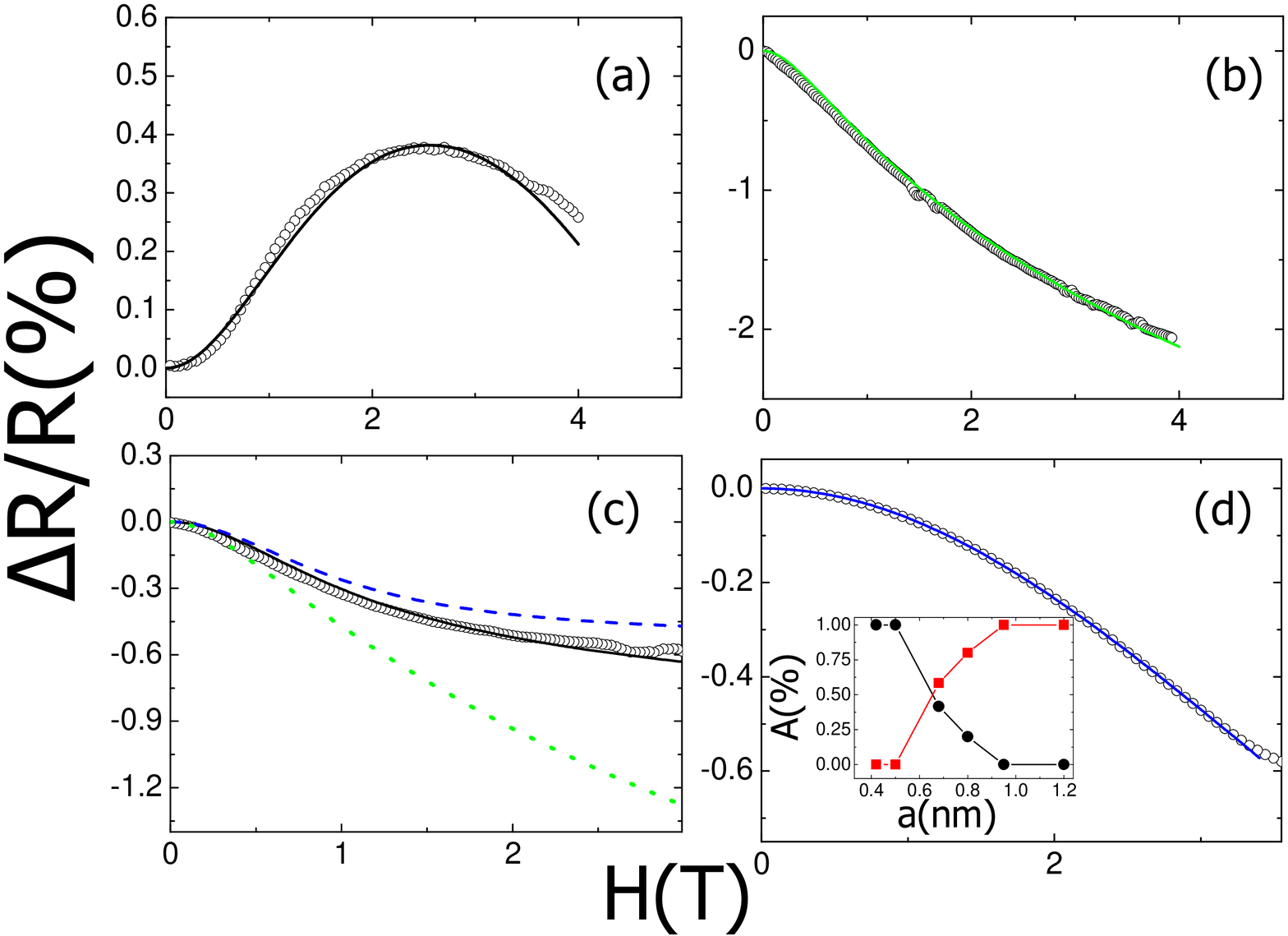}} \vspace{0.5CM} \caption{
\small MR curves of a Fe sample having various thickness $a$. T=4.2
K. (a) $a=0.16$ nm. The solid line is a fit to WAL (Eq. \ref{wal}).
(b) $a=0.5$ nm. The solid line is a fit to WL (Eq. \ref{wl}). (c)
$a=0.8$ nm. The solid line is a fit to combination between WL (the
dotted line) and AMR (the dashed line) of Eq. \ref{com}. (d) $a=7$
nm. The solid line is a fit to AMR (Eq. \ref{AMR}). The inset show
the amplitudes of WL (circles) and AMR (squares) ($A_{WL}$ and
$A_{AMR}$ of Eq. \ref{com})  as a function of the thickness.
 }\label{fe}
\end{figure}

It turns out that this crossover from ferromagnetic to non
ferromagnetic behavior has a striking effect on the
magnetoresistance. At $a=a_C$ the MR changes sign from positive to
negative. Fig. 3 depicts MR curves for a number of growth stages of
a Fe film, one for $a<a_C$ and the rest for $a>a_C$.  In the first
stage the MR is positive and is very similar to the results obtained
for Ag films having the same thickness. Here we obtain good fits to
WAL (solid line). As $a$ becomes larger than $a_C$ and the films are
characterized by spontaneous magnetization, the MR changes sign to negative, unlike the situation in Ag. For the
first few monolayers of thickness above $a_C$ the MR can be well
fitted to WL without the influence of SO in accoradance with Eq.
\ref{wl} (fig 3b).

\begin{figure}
{\epsfxsize=3.6 in \epsffile{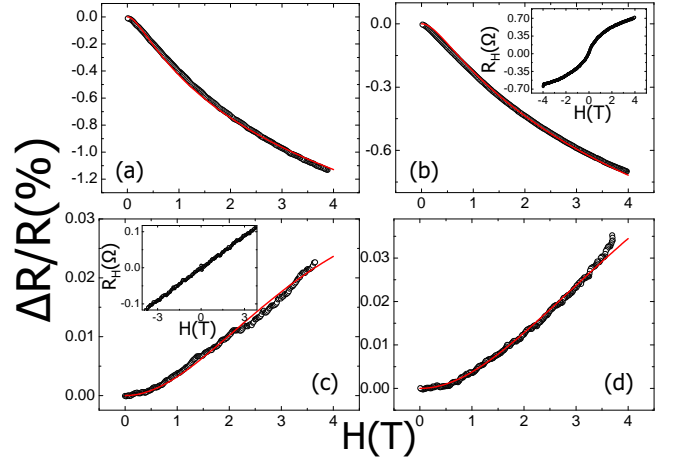}} \vspace{0.5cm} \caption{
\small MR curves of a 1.8 nm Ni film measured at the following
temperatures: (a) 0.5 K (b) 1.7 K  (c) 14 K, (d) 18 K. Solid lines
in (a) and (b) are fits to Eq. \ref{com}.  Solid lines in (c) and
(d) are fits to WAL (Eq. \ref{wal}). $l_{\varphi}$ extracted from
the fits are: 61.9, 48.9, 17.8 and 17.4 nm respectively. The insets
in (b) and (c) are HE measurements.
 }\label{temp}
\end{figure}

As the film is thickened
the curves can not be fitted by assuming WL contribution alone any more. For thick enough ferromagnetic films
we found \cite{our} that the MR curves are best described by anisotropic magnetoresistance (AMR) characteristic of
ferromagnetic films. The AMR effect relates the resistance value to
the angle, $\theta$, between the the current and the magnetization
\cite{AMR}. The dependance of the resistance on  $\theta$ is known
to be: $\rho\propto\cos^{2}\theta$, where the resistance is maximum
when the magnetization is parallel to the current. In our system the
applied magnetic field is always directed perpendicular to the
current. At zero field, both the magnetization and the current are
in the sample plane. As the applied field is increased, the
magnetization is rotated out of the plane thus increasing $\theta$.
At the saturation field, $H_{S}$, the magnetization is parallel to
the applied field and perpendicular to the current causing the
resistivity to be minimal. In order to analyze our data we assume
quadratic dependance of $\Delta$R on $H$ at fields lower than
$H_{S}$ and saturation value at high fields, and use the following
phenomenological expression:
\begin{equation}
 \label{AMR}   \Delta R_{AMR}=\Delta R(\infty)\frac{H^{2}}{H^{2}+H_{S}^{2}}
\end{equation}

The solid line in fig 3d is a fit to this expression where $H_s$ is
taken from the Hall effect measurements such as those of figure 2.
Hence, it appears that for thick enough ferromagnetic films the large
contribution of AMR overshadows all of other MR effects such as that
of weak localization.

In the intermediate regime (fig
3c), we fit the data to the combination between the two effects:
\begin{equation}
 \label{com}   \Delta R_{com}=A_{AMR}\Delta R_{AMR}+A_{WL}\Delta R_{WL}
\end{equation}
Where $A_{AMR}$ and $A_{WL}$ are coefficients determining the
relative weight of the two effects and $R_{WL}=1/\sigma_{WL}$. The
inset in fig. 3 which depicts the dependence of $A_{AMR}$ and
$A_{WL}$ on the thickness demonstrates that, for low thickness, the
negative MR observed in the ferromagnetic state is mainly due to WL
effects. As the film is thickened the relative AMR contribution
grows and becomes the dominant factor for large thickness.

The above experimental results are consistent with the theoretical
prediction of Dugaev et al \cite{dugaev}. For non-magnetic films,
the MR curves are always positive, presumably as a result  of the
corrections of SO interactions on WL.  For ferromagnetic films the
MR is always negative. The analysis of the data show that the
appearance of ferromagnetism in the film suppresses the effect of SO
such that only the usual negative MR typical to WL is present.

Fig 3 demonstrated a transition from paramagnetic to ferromagnetic
behavior as a function of thickness at constant temperature. A
similar effect can be observed at constant thickness as a function
of temperature.  This is shown in fig. 4 which depicts MR and HE
measurements of a 1.8 nm thick Ni film at different temperatures.
For $T<4$ $K$ (4a and 4b) the samples are ferromagnetic, as
demonstrated from the observed EHE, and the MR is negative. In this
regime the data fit a combination of WL and AMR of Eq. \ref{com}.
For $T>4$ $K$ (4c and 4d) only the ordinary HE can be observed,
indicating that the film has lost its ferromagnetism. Indeed, in
this regime the MR is positive and fits WAL behavior dominated by
SO.

\begin{figure}
{\epsfxsize=3.6 in \epsffile{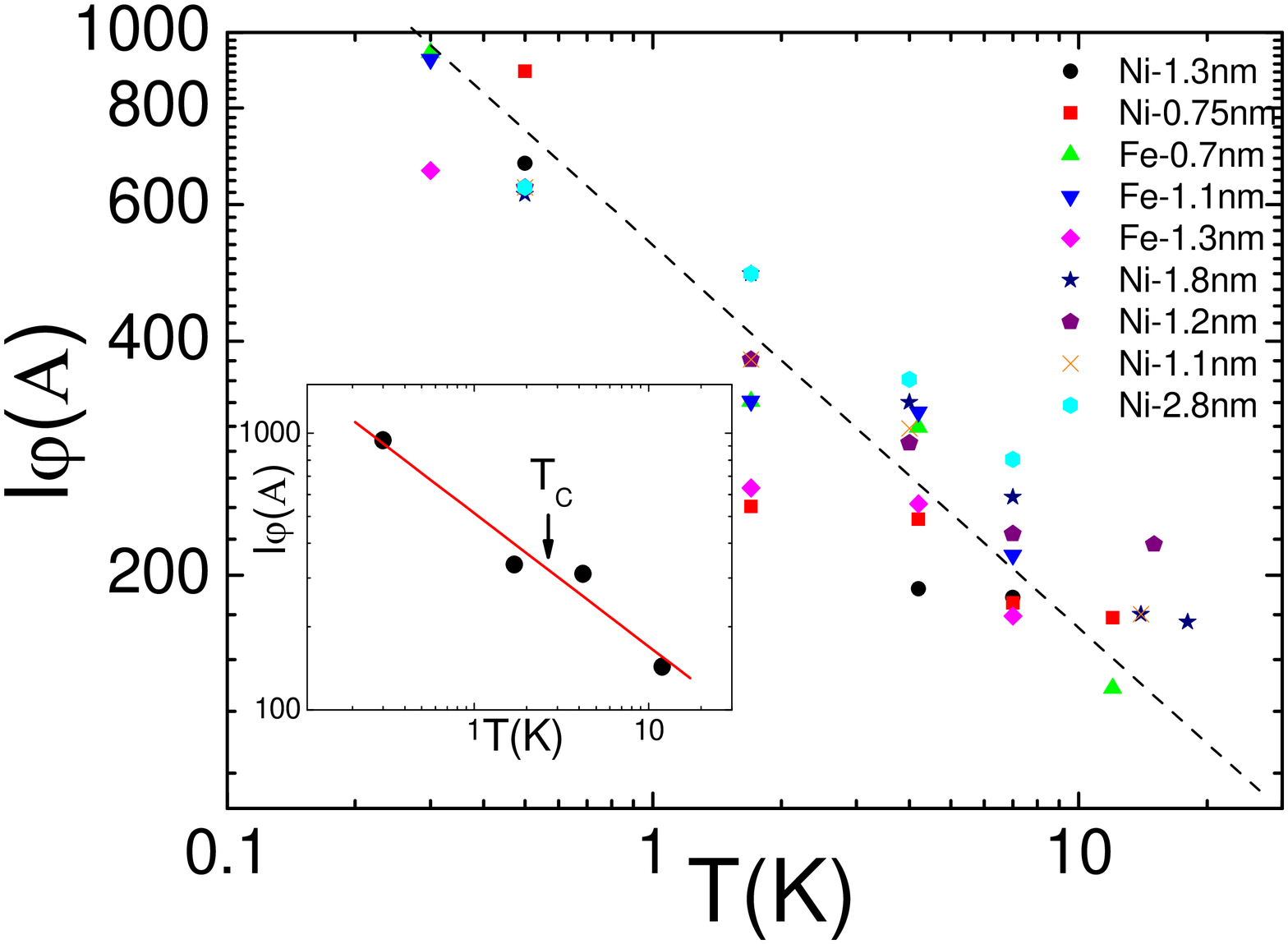}} \caption{ \small Dephasing
length, $l_{\varphi}$ as a function of T for a number of Ni and Fe
samples having different thicknesses.  The dashed line has a slope
of -0.5 and is a guide to the eye.  The inset is an example of one
Ni sample of 0.7 nm thick exhibits $l_{\varphi} \sim T^{-0.5}$.}
\label{lphi-T}
\end{figure}

Fits to WAL and WL similar to those of fig. \ref{temp} enable to
extract the values for the dephasing length, $l_{\varphi}$, at
different temperatures. It turns out that these values are very
similar for Ni and Fe. Fig. 5 shows $l_{\varphi}$ versus T for 9
samples of Ni and Fe having different thicknesses. All films exhibit
$l_{\varphi} \sim T^{-\alpha}$, with $\alpha$ ranging between
0.3-0.5. This power law is maintained even when the sample crosses
over from ferromagnetism to paramagnetism and the MR changes sign.
This is demonstrated in the inset which presents a 0.7 nm Ni film in
which $\alpha\sim0.5$. The crossover from WL to WAL, noted by the
arrow in the figure, is not sensed by $l_{\varphi}$ which shows a
smooth dependence on temperature.  This behavior reinforces our
confidence in the fitting procedure.

\begin{figure}
{\epsfxsize=3.2 in \epsffile{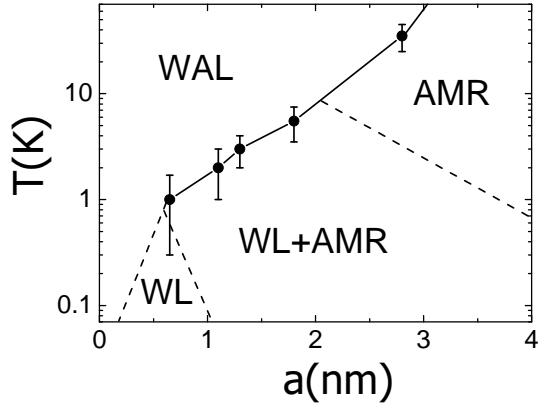}} \caption{ \small A schematic
illustration of the different MR regimes. The circles are the
measured Curie temperatures, $T_C$ as a function of thickness for a
typical Ni film. The solid line connecting them separates between
positive MR and negative MR. The dashed lines represent the
estimated qualitative separation between regimes of different
contribution to the negative MR} \label{T-d}
\end{figure}

\section{ \textrm{IV. Conclusions}}

In summary, we studied experimentally the magnetoresistance of
ultrathin films of amorphous Ni and Fe. The Curie temperature for
the ultrathin films of ferromagnetic metals decreases with the
decrease of the film thickness, and can be arbitrary low or even
zero making it possible to study both the paramagnetic state and the
ferromagnetic state in the same sample, at low enough temperatures
so that localization effects are prominent. The main results of this
paper are illustrated in fig. \ref{T-d} in which a schematic "phase
diagram" is plotted.  These can be summarized by the following
findings: At temperatures larger than the curie temperature, in
which the system shows no spontaneous magnetization, the MR is
positive and is attributed to weak antilocalization. This is
interpreted as signs for the fact that in the paramagnetic phase the
magnetoresistance is determined by the magnetic field dependence of
the quantum corrections to the conductivity in the presence of
strong Bychkov-Rashba SO scattering. This corresponds to the WAL
region  on the ``phase diagram''. At temperatures smaller than
$T_C$, the film becomes ferromagnetic and, at the same time, the MR
curves change sign and becomes negative. For low enough temperature
or thickness the MR curves follow the simple weak localization
behavior without SO scattering (corresponding to the WL regime of
fig. \ref{T-d}).  This indicates that scattering in the triplet
channel, leading to WAL in the presence of SO scattering, is
suppressed, and only scattering in the singlet channel is effective.
As the film is thickened or temperature raised the magnetoresistance
due to quantum correction to the conductivity is masked partially or
completely  by the anisotropic magnetoresistance effects. This
corresponds to WL+AMR and AMR regions on the  ``phase diagram''
respectively.

Thus we have found that ultrathin films of amorphous Ni and Fe, interesting by itself due
to possible applications, can serve as a testing ground for the theory
of quantum corrections to the conductivity. The experimental results agree qualitatively with the theoretical prediction \cite{dugaev} that ferromagnetism in a film totally suppresses
the influence of spin orbit scattering on the perpendicular magnetoresistance. Though there is no theory discussing the interplay between the different contributions of WL and AMR to the magnetoresistance curve in such films, the analysis of the experimental results provides information about the smooth transition from the WL dominated regime to the region in which AMR governs the behavior.

We are grateful for useful discussions with L. Klein. This research was supported by the Israeli academy of science (grant number 399/09)

\end{document}